\documentclass[12pt]{article}

\usepackage[cp1251]{inputenc}
\usepackage[centertags]{amsmath}
\usepackage{amsfonts}
\usepackage{amssymb}
\usepackage{amsthm}
\usepackage{amsmath}
\usepackage{newlfont}
\usepackage{graphicx}

\date{}
\title{Quarkonia potential}
\author{ M. A. Durnev \\  {\em Department of Theoretical Physics }\\ 
 {\em St. Petersburg State University, 198904,} \\
 {\em St. Petersburg, Russia}\\
 {\em E-mail: mad5245@mail.ru}\\} 
\scrollmode

\begin{document}
\makeatletter
\renewcommand{\@oddhead}{\hfil {\em Quarkonia potential}}
\makeatother
\maketitle

\begin{abstract}
{\small 
\noindent Using the quark-antiquark interactions obtained in the framework of the bootstrap method we
construct a potential model, investigate the possibility of describing of heavy quarkonia and 
calculate the bottomonium spectrum. The potential of the interaction was obtained as a
nonrelativistic limit of the relativistic quark-antiquark amplitudes $Q{\bar Q} \to Q\bar Q$. }
\end{abstract}
{\bf 1. Introduction \\}
Quarkonium systems, the bound states of a heavy quark and antiquark, have played a particularly 
important role in considering of strong interaction dynamics. The discovery of heavy quarkonia: 
the families of $J/\psi$- and $\Upsilon$-mesons promoted the quark model and non-Abelian gauge 
field theories, leading to the prevailing picture of particle physics.
\\
QCD features two remarkable properties. First, asymptotic freedom implies that at very high 
energies and momenta, quarks and gluons interact only weakly and act as quasifree particles [1, 2].
Second, confinement presumably results from the fact that at low energies the force between
quarks increases with their distance, so that quarks are always tied into hadrons and cannot
be removed individually.
\\
Confinement makes it hard to calculate quantities for the bound states within QCD as one cannot
apply perturbative QCD. By analogy with the positronium, and given the large masses of the charm and
bottom quarks, nonrelativistic phenomenological potential models have been applied as tools
for the quarkonium spectroscopy. To accommodate the properties of QCD these models, {\em e.g.} [3-6], 
are based on a short range part motivated by perturbative QCD and a phenomenological long range 
part accounting for confinement.
\\
In QCD with heavy $c-$ and $b-$quarks, the characteristic scale $\Lambda_{QCD} \sim$ 0.2 GeV is small as
compared to the quark masses, $m_{c} \sim$ 1.5 GeV and $m_{b}\sim$ 5 GeV. A systematic expansion in the powers
of $1/m_{Q}$ is possible [7-9]. The bound state problem can be dealt with non-relativistically. Following
these thoughts one treats the bound state problem by solving a Schroedinger equation using an
appropriate potential.
\\
In this work the quark interaction potential is considered. The short range part of the potential 
is obtained as a nonrelativistic limit of the relativistic quark amplitudes of the bootstrap quark model [10, 11].
These quark amplitudes depend not only on a squared  momentum transfer $t$, but on the energy variable
$s$ also. Therefore the direct transition to nonrelativistic potentials is not possible: these amplitudes
correspond more to quasipotentials [12]. To obtain quark potentials from the quark amplitudes one has 
to fix the energy $s=s_{0}$ and then the dependence on momentum tranfer at fixed energy is considered 
to be potential. The energy fixing $s$ implies an introduction of a momentum cutoff parameter 
$\Lambda_{F}$ in the Fourier transformation. As a result, 
the following expression is obtained for the short range part of the potential [13]:
{\multlinegap=0pt \begin{multline}
V_{B}(r)=- \dfrac{1} {m_{q}^2}\int\limits_{0}^{\Lambda_{\Phi}} \dfrac {k} {r} \, \sin{kr}\dfrac{g}
{1-gB(k^2)}dk, \qquad \qquad \qquad \qquad \qquad \qquad \quad \tag{1} \end{multline} } 
where $g$ is a dimensionless coupling constant, which is also a parameter of the model,
$B(k^2)$ is the Chew-Mandelstam function [14] for the gluon state:\\ 
 {\multlinegap=0pt \begin{multline}
 B(k^{2})=\biggl(-\beta_{1} \dfrac {k^{2}} {4m^{2}}+\beta_{2} \biggr) \sqrt{\dfrac {k^{2}+4m^{2}} {k^{2}}} \;
 \ln {\dfrac{\sqrt{\dfrac{k^{2}+4m^{2}} {k^{2}}}+\sqrt{\dfrac {\Lambda-4m^{2}} {\Lambda}}}
 {\sqrt{\dfrac {k^{2}+4m^{2}} {k^{2}}}-\sqrt{\dfrac {\Lambda-4m^{2}} {\Lambda}}}} \;+ \\[12pt] 
 +\; \beta_{1} \dfrac{\sqrt{\Lambda(\Lambda-4m^{2})}}
 {4m^{2}} \; + \biggl(\beta_{2}-\beta_{1} \biggl(\dfrac {k^{2}} {4m^{2}} + \dfrac {1} {2} \biggr)\biggr)
 \ln {\dfrac {1+\sqrt{\dfrac{\Lambda-4m^{2}} {\Lambda}}}{1-\sqrt{\dfrac{\Lambda-4m^{2}} {\Lambda}}}} \; ,
 \qquad \qquad \quad \quad \raisetag{70pt} \tag{2} \end{multline} }\\
 where $m$ is the mass of a heavy quark $m_{b}$, the coefficients $\beta_{1}$ and $\beta_{2}$  for $1^{-}$ state are:
 $\beta_{1}=\dfrac{1} {3}, 
 \beta_{2}=\dfrac {1} {6}$.\\
The qualitative behaviour of the bootstrap potential with $r$ is shown in {\em fig.}1.
\\
In difference with the majority of the quark interaction potentials [5, 6, 15-20] the bootstrap potential 
has the finite value at $r=0$  in the consequence of energy cutoff introduced in calculations of bootstrap
quark amplitudes. Besides this, the presence of the momentum cutoff $\Lambda_{F}$ leads to
the origin of small oscillations of the potential at distances $\sim$ 1 fm.
\\
The potential of confinement is considered as a linear potential with a slope defined by the
angle $\alpha$. This potential is added to the bootstrap potential at a distance $r_{0}$.
$\alpha$ and $r_{0}$ are also the parameters of our potential model. So the quarkonia potential
has the form something like that shown in {\em fig.}2.\\
\\
{\bf 2. Results}
\\
\\
The quarkonia potential considered in the previous section is used in the time-independent Schroedinger equation 
to find bound states, while spin-spin and spin-orbit interactions (Breit-Fermi interaction) are treated 
within perturbation theory. The resulting mass formula is given by
 {\multlinegap=0pt \begin{multline}
M(k^{2S+1}l_{j})=2m_{Q}+E_{kl}+\dfrac{32\pi\alpha_{s}} {9m_{Q}^2} \biggl(\dfrac{1} {2}S(S+1)-
\dfrac{3}{4} \biggr)|\psi_{kl}(0)|^2+ \\ \qquad \qquad \qquad \qquad \qquad \qquad \qquad  
+\alpha_{s}\dfrac{j(j+1)-l(l+1)-S(S+1)} {m_{Q}^2}
\biggl<\dfrac{1} {r^3}\biggr>, 
\qquad   \raisetag{45pt} \tag{3} \end{multline} }
where $\alpha_{s}$ is a running coupling constant.

We use as input to determine the model parameters the states  $\eta_{b}(1S)$, $\Upsilon(1S)$, $\Upsilon(2S)$ and
$C(1P)$, the center of gravity for the $1P$ triplet states defined as
\\
$C(1P) = \dfrac {1} {9} (5M(\chi_{b2})+3M(\chi_{b1})+M(\chi_{b0})) \approx 9900$ MeV. \qquad \qquad \qquad \qquad  (4)
\\
Identifying the $C(1P)$ with the $1^{1}P_{1}$ state of the model the resulting parameter set is \\
$\Lambda_{F}=3.05$ GeV,\\
$g=2.22$,\\
$\alpha= 0.28,$ \qquad \qquad \qquad \qquad \qquad \qquad \qquad \qquad \qquad \qquad \qquad \qquad \qquad \qquad (5)\\
$r_{0}=0.39$ fm,\\
the values of  $\Lambda_{b}$ (in the $B$-function) and $m_{b}$ are taken from the bootstrap method [21].
\\
The resulting masses are compared with the experimental data in {\em table} 1, while in {\em fig.}3 the resulting
spectrum is displayed together with the experimental one. The bottomonium S-wave state reduced radial wavefunctions
calculated within the model are shown in {\em fig.}4 and for the P- and D-wave states are shown in {\em fig.}5a,b.

\newpage
\noindent {\bf 3. Conclusion}\\ 
\\
In this work, we have costructed a potential model for heavy quarkonium based on
the bootstrap potential. The aim of this work has been to investigate the possibility of
describing of heavy quark-antiquark systems with the help of the derived potential
using the Schroedinger equation and to get a satisfactory description of the quarkonium
spectra with minimal phenomenological input.
\\
We have demonstrated that a satisfactory description of the quarkonium spectra is possible
within this model with reasonable values of the parameters. In the future one should use it
to calculate other properties such as quarkonia radii, decay widths and branching ratios.\\
\\
\\
\\
\\
{\bf 4. Acknowledgments}\\ \\
The author is grateful to S. M. Gerasyuta for support and discussions.
This research is supported in part by Russian Ministry of Education (grant 2.1.1.68.26).

\newpage
{\bf References}\\ 
1. D.J. Gross, F. Wilczek, Phys. Rev. Lett. {\bf 30}, 1343 (1973).\\
2. H.D. Politzer, Phys. Rev. Lett.{\bf 30}, 1346 (1973).\\
3. A.De Rujula, H. Georgi, S.L. Glashow, Phys. Rev. D {\bf 12}, 147  

(1975).\\
4. E. Eichten, K. Gottfried, T. Kinoshita, K.D. Lane, T.-M. Yan, 

Phys. Rev. D {\bf 17}, 3090 (1978).\\
5. J.L. Richardson, Phys. Lett. B {\bf 82}, 272 (1979).\\
6. W. Buchmuller, S.H.H. Tye, Phys. Rev. D {\bf 24}, 132 (1981).\\
7. N. Isgur, M.B. Wise, Phys. Lett. B {\bf 232}, 113 (1989).\\
8. H. Georgi, Phys. Lett. B {\bf 240}, 447 (1990).\\
9. A.F. Falk, H. Georgi, B. Grinstein, M.B. Wise, Nucl. Phys. B {\bf 343}, 

1 (1990).\\ 
10. V.V. Anisovich, S.M. Gerasyuta, Sov. J. Nucl. Phys. {\bf 44}, 174 (1986).\\
11. V.V. Anisovich, S.M. Gerasyuta, A.V. Sarantsev, Int. J. Mod. Phys. 

{\bf 6}, 625 (1991).\\
12. A.A. Logunov, A.N. Tavkhelidze, Nuovo Cim. A {\bf 29}, 370 (1963).\\
13. S.M. Gerasyuta, Yu.A. Kuperin, A.V. Sarantsev, E.A. Yarevsky, 

Yad. Fiz. {\bf 53}, 1397 (1991).\\
14. G.F. Chew, {\em The Analytic S-Matrix}, (Benjamin, New York, 1966). \\
15. R.H. Dalitz, Nucl. Phys. A {\bf 353}, 251 (1981).\\
16. S. Ono, Preprint CERN TH.3679, 1983.\\
17. S.N. Gupta, S.F. Radford, W.W. Repko, Phys. Rev. D {\bf 26}, 3305 (1982).\\
18. I.M. Dremin, A.V. Leonidov, TMF {\bf 51}, 178 (1982).\\
19. A.J.D. Hey, R. Kelly, Phys. Rep. C {\bf 96}, 71 (1983).\\
20. A. Martin, Preprint CERN TH-4060/84, 1984; Preprint CERN 

TH-6604/85, 1985.\\
21. S.M. Gerasyuta, A.V. Sarantsev, Yad. Fiz. {\bf 52}, 1483 (1990).\\
22. W.-M. Yao {\em et al.,} J. Phys. G {\bf 33}, 1 (2006).\\
23. J.L. Rosner, hep-ph/0606166.

 \newpage
 \noindent
 {\bf Figure and table captions} \\ \\
 Figure 1: The bootstrap potential as a short range part of the model potential.\\ \\
 Figure 2: The model potential.\\ \\
 Figure 3: Bottomonium spectra of experiment and our model; $\eta_{b}(2S)$, $\eta_{b}(3S)$,
  $h_{b}(1P)$ and $h_{b}(2P)$ mass measurements [23] are included separately in the experimental spectrum.\\ \\
 Figure 4: Bottomonium S-wave state reduced radial wavefunctions calculated within our model.\\ \\
 Figure 5a: Bottomonium P-wave state reduced radial wavefunctions calculated within our model.\\ \\
 Figure 5b: Bottomonium D-wave state reduced radial wavefunctions calculated within our model. \\  \\
 Table 1: $b\bar b$-state masses from the experiment and our model.

 \newpage
 \includegraphics[scale=0.5]{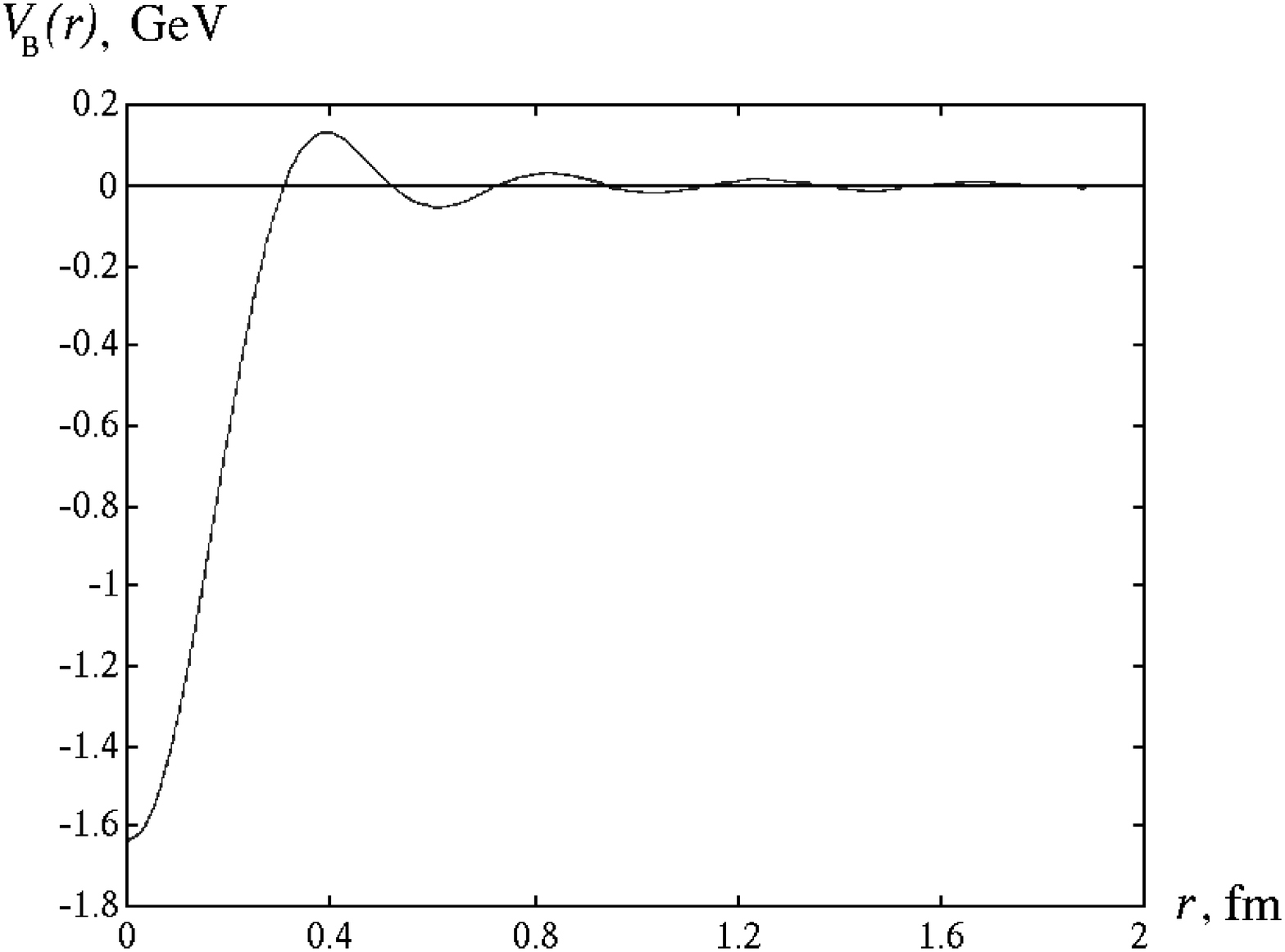}\\
 \newpage
 \includegraphics[scale=0.5]{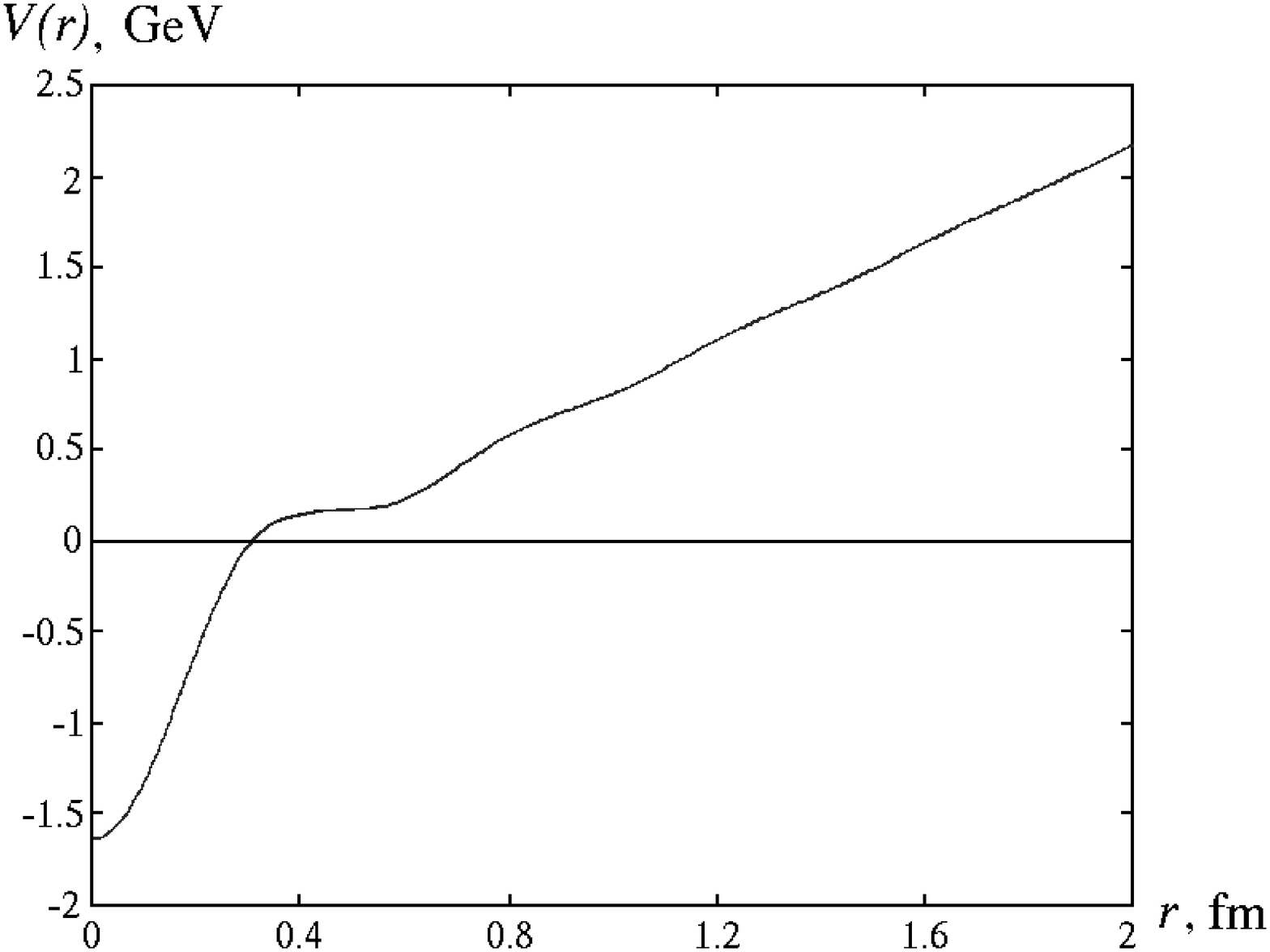}\\
  \newpage
 \includegraphics[scale=0.5]{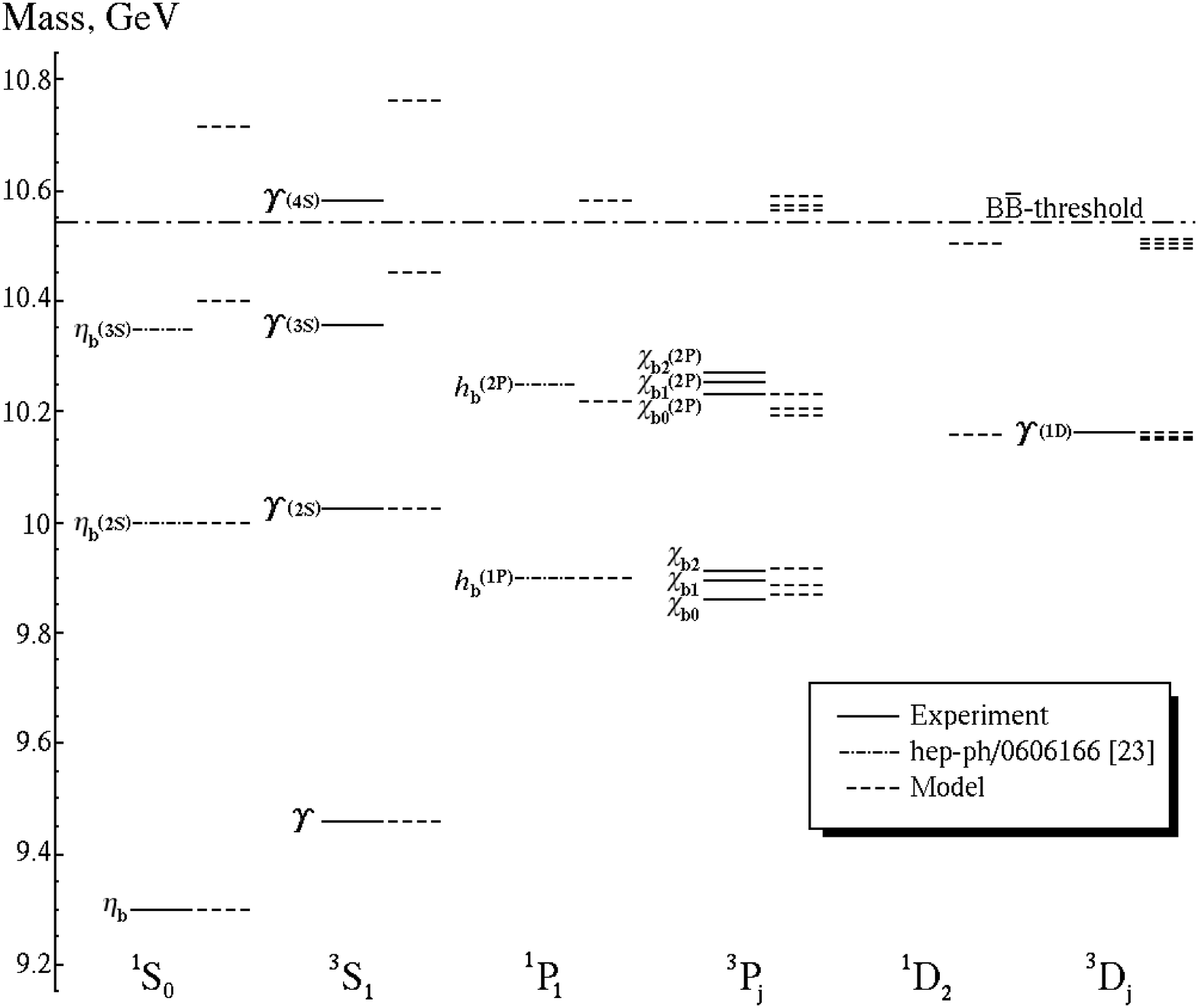}\\
 \newpage
 \includegraphics[scale=0.5]{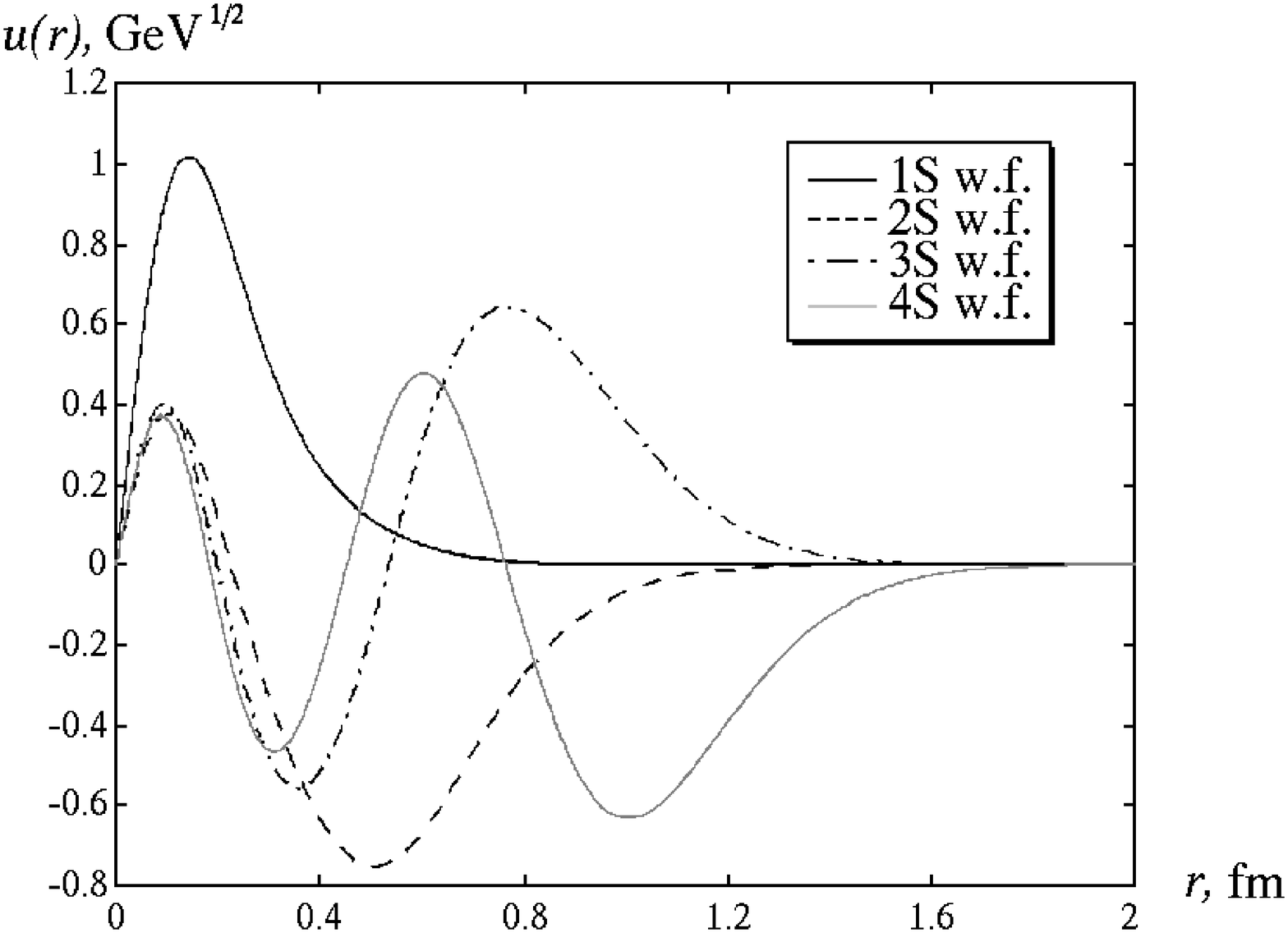}\\
 \newpage
 \includegraphics[scale=0.5]{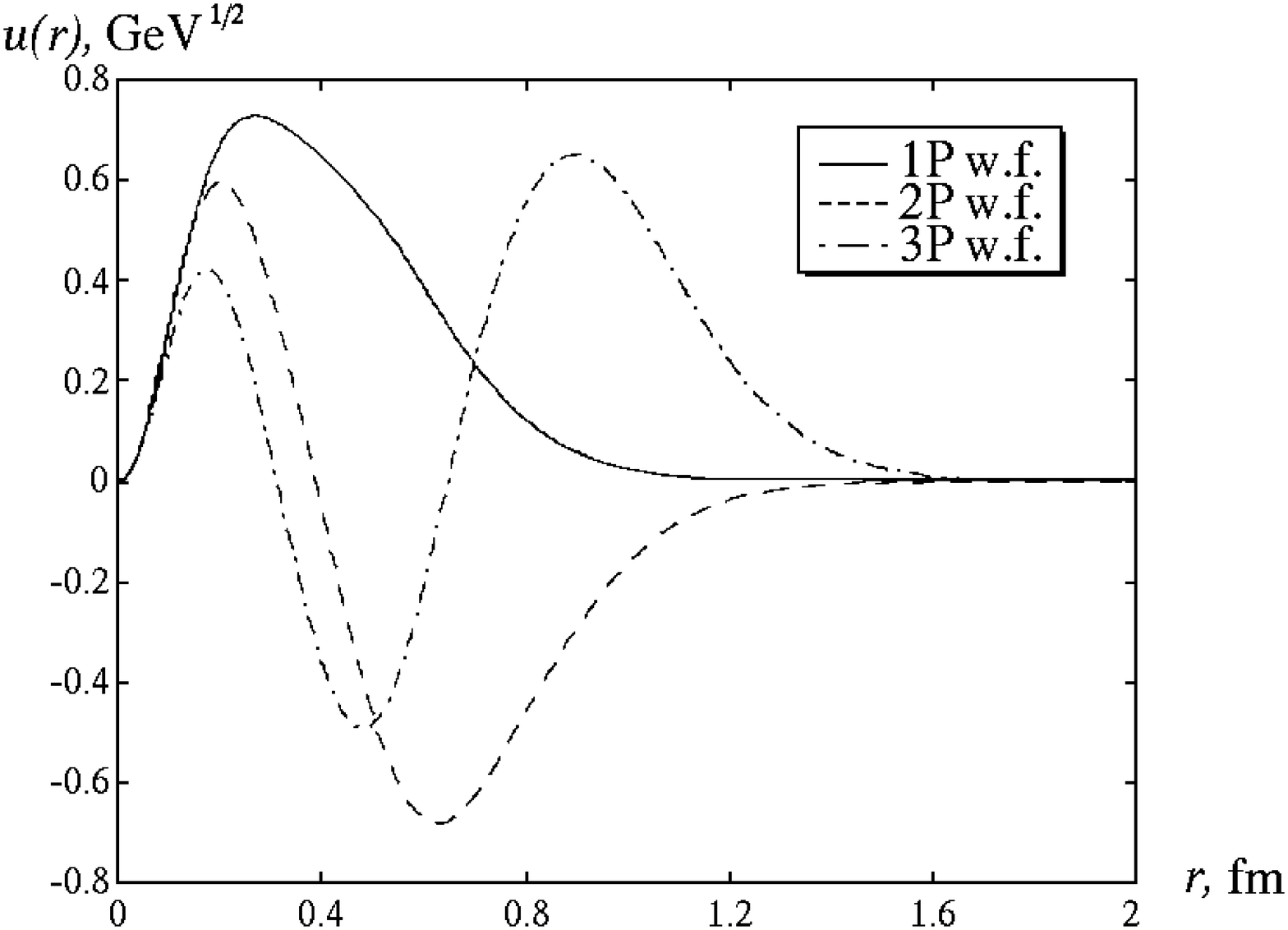}\\
 \newpage
 \includegraphics[scale=0.5]{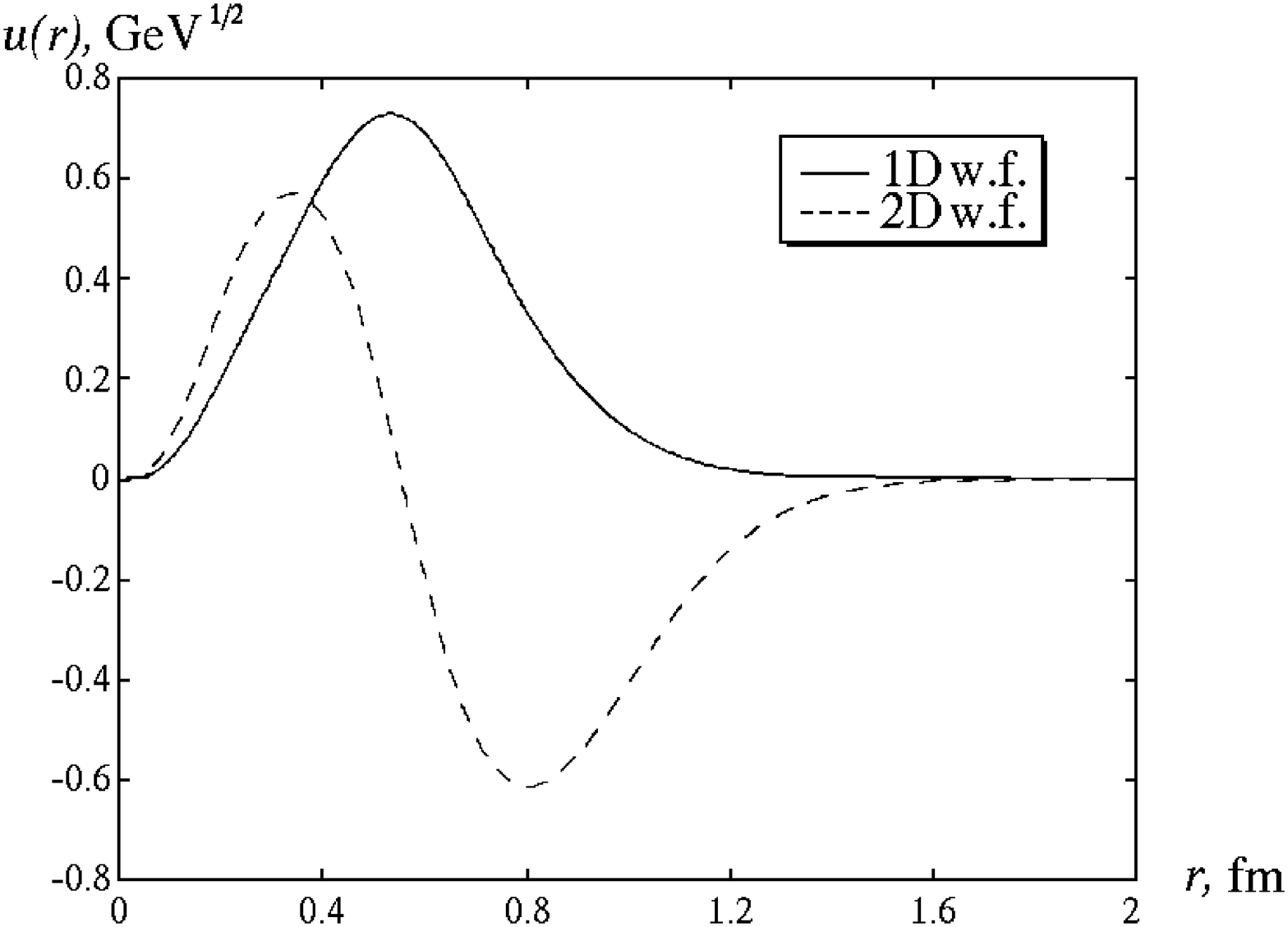}\\

\newpage
\noindent Table 1. $b\bar b$-state masses from the experiment and our model.
\\[8pt]  
\begin{tabular} {|c|c|c|c|}
\hline
State & Candidate  & Experimental mass [22], MeV & Theoretical mass, MeV    \\
\hline
$1^{1}S_{0}$ & $\eta_{b}(1S)$ & 9300 $\pm$ 20 $\pm$ 20 & 9300\\[-5pt]
$1^{3}S_{1}$ & $\Upsilon(1S)$ & 9460.30 $\pm$ 0.26 & 9460 \\[-3pt]
\hline
$1^{1}P_{1}$ & \qquad  & \qquad & 9900 \\[-5pt]
$1^{3}P_{0}$ &  $\chi_{b0}(1P)$ & 9859.44 $\pm$ 0.42 $\pm$ 0.31 & 9869\\[-5pt]
$1^{3}P_{1}$ &  $\chi_{b1}(1P)$ & 9892.78 $\pm$ 0.26 $\pm$ 0.31 & 9884\\[-5pt]
$1^{3}P_{2}$ & $\chi_{b2}(1P)$ & 9912.21 $\pm$ 0.26 $\pm$ 0.31 & 9916\\[-3pt]
\hline
$2^{1}S_{0}$ & \qquad & \qquad & 9997\\[-5pt]
$2^{3}S_{1}$ & $\Upsilon(2S)$ & 10023.26 $\pm$ 0.31 & 10023 \\[-3pt]
\hline
$1^{1}D_{2} $ & \qquad & \qquad & 10156\\[-5pt]
$1^{3}D_{1} $ & \qquad & \qquad & 10150\\[-5pt]
$1^{3}D_{2} $ & $\Upsilon(1D)$ & 10161.1 $\pm 0.6 \pm$ 1.6 & 10154 \\[-5pt]
$1^{3}D_{3} $ & \qquad & \qquad & 10160\\[-3pt]
\hline
$2^{1}P_{1} $ &\qquad & \qquad & 10219\\[-5pt]
$2^{3}P_{0} $ & $\chi_{b0}(2P)$ & $10232.5 \pm 0.4 \pm 0.5 $ & 10191 \\[-5pt]
$2^{3}P_{1} $ & $\chi_{b1}(2P)$ & $10255.46 \pm 0.22 \pm 0.50$ & 10205 \\[-5pt]
$2^{3}P_{2} $ & $\chi_{b2}(2P)$ & $10268.65 \pm 0.22 \pm 0.50$ & 10233 \\[-3pt]
\hline
$3^{1}S_{0}$ & \qquad & \qquad & 10400\\[-5pt]
$3^{3}S_{1}$ & $\Upsilon(3S)$ & $10355.2 \pm 0.5 $ & 10450 \\[-3pt]
\hline
$2^{1}D_{2} $ & \qquad & \qquad & 10505\\[-5pt]
$2^{3}D_{1} $ &\qquad & \qquad & 10495\\[-5pt]
$2^{3}D_{2} $ &\qquad & \qquad & 10502 \\[-5pt]
$2^{3}D_{3} $ &\qquad & \qquad & 10511\\[-3pt]
\hline
$3^{1}P_{1} $ &\qquad & \qquad & 10580\\[-5pt]
$3^{3}P_{0} $ &\qquad & \qquad & 10562\\[-5pt]
$3^{3}P_{1} $ &\qquad & \qquad & 10571\\[-5pt]
$3^{3}P_{2} $ &\qquad & \qquad & 10589\\[-3pt]
\hline
$4^{1}S_{0} $ &\qquad & \qquad & 10716 \\[-5pt]
$4^{3}S_{1} $ & $\Upsilon(4S) $ & 10579.4 $\pm 1.2 $ & 10764\\
\hline
\end{tabular}\\

\end{document}